\documentclass[aps,prb,twocolumn,showpacs]{revtex4-1}
\usepackage{bm}
\usepackage{graphicx}
\usepackage{cancel}
\usepackage{braket}
\usepackage{mathrsfs}
\usepackage{mathtools}
\usepackage{tabularx}
\usepackage{multirow}

\begin{document}

\title{Accurate ionization potential of semiconductors from efficient density functional calculations}

\author{Lin-Hui Ye}
\affiliation{Key Laboratory for the Physics and Chemistry of Nanodevices, Department of Electronics, Peking University, Beijing 100871, P.R. China}

\date{\today}

\begin{abstract}
Despite of its huge successes in total energy related applications, the Kohn-Sham scheme of density functional theory (DFT-Kohn-Sham) cannot get reliable single particle excitation energies for solids. In particular, it has not been able to calculate the ionization potential (IP), one of the most important material parameters, for semiconductors. We illustrate that an approximate exchange-only optimized effective potential (EXX-OEP), the Becke-Johnson exchange, can be used to largely solve this long-standing problem. For a group of 17 semiconductors, we have obtained the IP's to similar accuracy as the much more sophisticated $GW$ approximation (GWA), with the computational cost of only LDA/GGA. The EXX-OEP, therefore, is likely as useful for solids as for finite systems.  For solid surfaces, the asymptotic behavior of the $v_{xc}$ has similar effects as for finite systems which, when neglected, typically causes the semiconductor IP's to be underestimated by about 0.2 eV. This may partially explain why standard GWA systematically underestimates the IP's, and why using the same GWA procedures has not been able to get accurate IP and band gap at the same time. 
\end{abstract}

\pacs{71.15.Mb, 77.22.Ej}

\maketitle

The ionization potential of a semiconductor is one of the most important material parameters which governs its optics, transport, electrochemistry and interface properties in hetero-structures, etc. Although in principle the IP can be calculated by the many-body Green's function method in $GW$ approximation (GWA)\cite{GWA}, due to the heavy computational cost, in practice further approximations have to be made which has remained unjustified to this date. For finite systems, it is usually more convenient to use density functional theory (DFT) to calculate the neutral and ionized systems separately, and then obtain accurate IP from their total energy difference. This approach, however, cannot be used for infinite solids for which DFT can only treat neutral systems.

Lucky enough,  in DFT there exists an ``IP theorem''\cite{theorem1,theorem2} which states that the negative eigenvalue of the highest occupied Kohn-Sham orbital, -$\varepsilon_N$, happens to be equal to the IP. However, the IP theorem relies on the exact exchange-correlation potential $v_{xc}$ which is unknown and must be approximated. Popular approximations such as the local density approximation (LDA) or generalized gradient approximations (GGA) suffer from the self-interaction (SI) error\cite{SIC} which pushes up occupied states and causes LDA/GGA to systematically underestimate the IP. For atoms and molecules\cite{LB94}, the IP error (Hereafter ``IP error'' specifically refers to the deviation of the  $-\varepsilon_{N}$ of LDA/GGA from the true IP) can be as large as  $5\sim 10$ eV, while for semiconductors\cite{Gorling,jiangip} it is typically 1 eV. Consequently, the powerful DFT-Kohn-Sham scheme has not been able to get reliable IP's for semiconductors even as simple as silicon.

In this paper we make a major step forward, showing that by use of the optimized effective potential\cite{OEP1,OEP2,OEP3} within the authentic slab geometry\cite{mypaper} the IP theorem can predict semiconductor IP\cite{ipnote} as accurate as the much more sophisticated GWA. Since the whole procedure is restricted to the DFT-Kohn-Sham realm, the computational cost is as low as LDA/GGA. For the first time, this opens the way for systematic calculation of semiconductor IP's by the DFT-Kohn-Sham scheme.

The condition for accurate IP is much more stringent than for accurate total energy, since $v_{xc}$ depends on every detail of the exchange-correlation hole while total energy only requires the spherical average of the hole to be correct. This explains why LDA/GGA behave drastically different for the two quantities.  For finite systems, it has long  been established\cite{LB94} that the exact $v_{xc}$ must decay as slowly as $-1/r$ away from the physical system. However, due to the SI error LDA/GGA decay exponentially fast. Consequently, a large negative portion of $v_{xc}$ is missing  which is responsible for most of the eigenvalue overestimation. Often, the accuracy of IP can be improved by correcting the wrong asymptotic behavior of the $v_{xc}$. However, for solid surfaces asymptotically long-ranged potentials are very hard to use with the supercell method\cite{mypaper}, so that the role of the long range part of $v_{xc}$ has remained unexplored. As revealed by this work, the effects are not as strong as for finite systems, but still too significant to be simply ignored.

The exact-exchange only optimized effective potential (EXX-OEP, or simply OEP as in this work) is the local multiplicative potential minimizing the Hartree-Fock expression of total energy. Using OEP to calculate semiconductor IP is inspired by the fact that most SI resides in the exchange part of $v_{xc}$, while OEP is self-interaction free\cite{MESI}, and satisfies many exact conditions especially the desired $-1/r$ asymptotic behavior\cite{KLI2}. For neutral and ionized atoms\cite{OEPip} as well as small molecules\cite{Gorling}, OEP removes the IP errors by more than 80\%.

The capability of OEP to calculate semiconductor IP is not trivially seen because its success for atoms and molecules seems to rely on the overwhelmingly dominant role of exchange, while in solids correlation is much stronger. In fact, so far there is only one work\cite{Engel} using OEP  to calculate semiconductor IP's. For graphene the obtained value is 8.13 eV which severely overestimates the experimental IP of 4.6 eV.  Consequently, the author concluded that the correlation potential for graphene has to be repulsive. For Si (111) the obtained value is 5.45 eV which still overestimates the experiment by 0.4 eV. In fact, The Si result is unreliable since the slab is too thin (It contains only three bilayers) and the surface is not relaxed. In general, OEP is not considered useful for semiconductor IP, a conclusion we will challenge in this paper.

The application of OEP for solids has not become popular mainly because the construction of the exact OEP is technically very challenging: OEP is computationally costly, and is extremely sensitive to subtle balance of the basis sets\cite{basis1,basis2,Betzinger}. For semiconductor IP the problem is agonized due to its extra surface dependence which therefore requires the OEP of the complex surface structure. Moreover, OEP decays as slowly as $-1/z$ from surface. Like other long-ranged potentials, it is very hard to use with the  supercell method\cite{Engel,mypaper}.

This work is made possible by implementing an approximate OEP, the Becke-Johnson'06 (BJ06) exchange potential\cite{BJ06}, to the authentic slab geometry (ASG) which is a surface technique suitable for both short- and long-ranged potentials\citep{mypaper}. Even for simple systems, using approximate OEP to avoid the intricacy of the exact OEP is common. For example, the Krieger-Li-Iafrate (KLI) approximation\cite{KLI1} has been much more popular than the exact OEP itself. Compared to KLI, BJ06 is less accurate. But for the following reasons BJ06 shall well represent the performance of OEP: $(i)$ For atoms BJ06 approaches the exact OEP very closely. $(ii)$ For solids it satisfies the uniform electron gas limit. $(iii)$ For finite systems BJ06 decays as $-1/r+C$ ($C$ is positive and system dependent), and it is found\cite{BJ06ip} to reduce the IP errors by as much as 60\%. $(iv)$ For solid surface BJ06 decays as\cite{mypaper} $-1/z+C$  which will be shown to be equivalent to the asymptotic behavior of the exact OEP. To account for the stronger correlation in solids, in this work BJ06 is amended by LDA correlation\cite{GGAnote}.  The total $v_{xc}$ is termed BJ06c throughout this paper.
\begin{figure}[h]
\centering
\includegraphics[width=\linewidth]{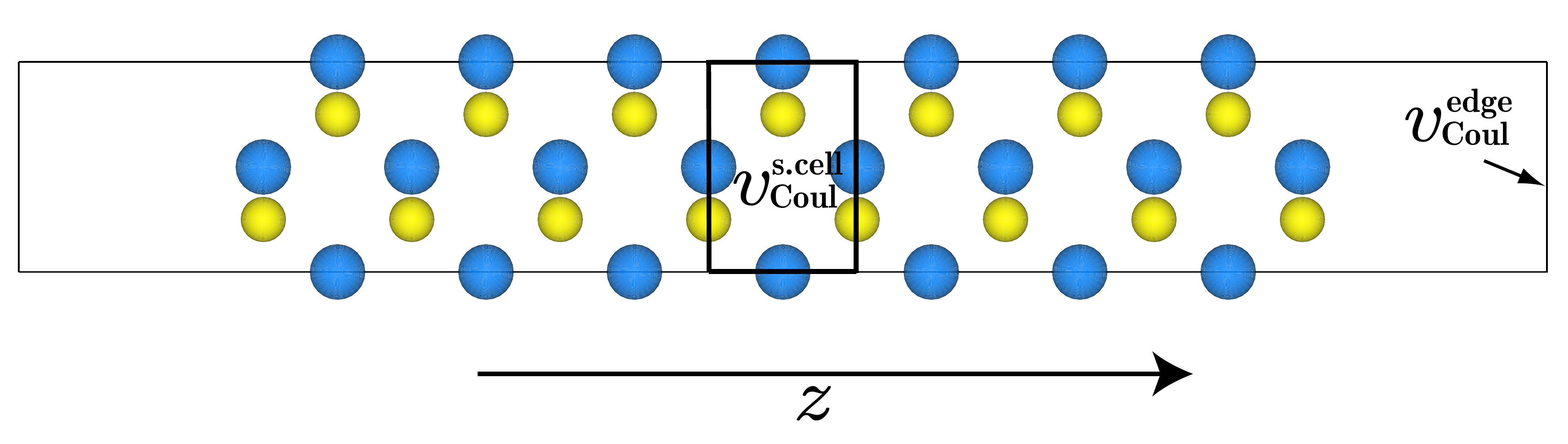}
\caption{\label{ZnSslab} (Color on line) A ZnS slab of 15 atom layers. The highlighted region is used for macroscopic averaging\cite{Fall}.}
\end{figure}

The IP calculation follows the usual two-step procedure: First by a bulk calculation (with the primitive cell) to get $\varepsilon_{\mathrm{vbm}}$, the eigenvalue of the valence band maximum (VBM); Then by a surface calculation using a slab like FIG.\ref{ZnSslab} for the alignment of the Coulomb potential. The surface calculation is needed since in the bulk calculation the Coulomb potential is determined only up to a global constant due to the use of periodic boundary conditions. The expression of IP used in this work is:
\begin{align}
\label{IPsolidFinal}
\mathrm{IP} = -\varepsilon_{\mathrm{vbm}}^{\mathrm{p.cell}} + \overline{v}_{\mathrm{Coul}}^{\mathrm{p.cell}}-\overline{v}_{\mathrm{Coul}}^{\mathrm{s.cell}},
\end{align}
in which $\overline{v}_{\mathrm{Coul}}^{\mathrm{p.cell}}$ is the average Coulomb potential of the bulk, and $\overline{v}_{\mathrm{Coul}}^{\mathrm{s.cell}}$ the drop of the average Coulomb potential from the vacuum edge to the center of the slab. Since $\overline{v}_{\mathrm{Coul}}^{\mathrm{s.cell}}$ is caused by the formation of surface dipoles, it is very sensitive to the surface charge distribution which is the first hint that the long-range part of the $v_{xc}$ shall play an important role in the final IP.

\begin{figure}[h]
\centering
\includegraphics[width=\linewidth]{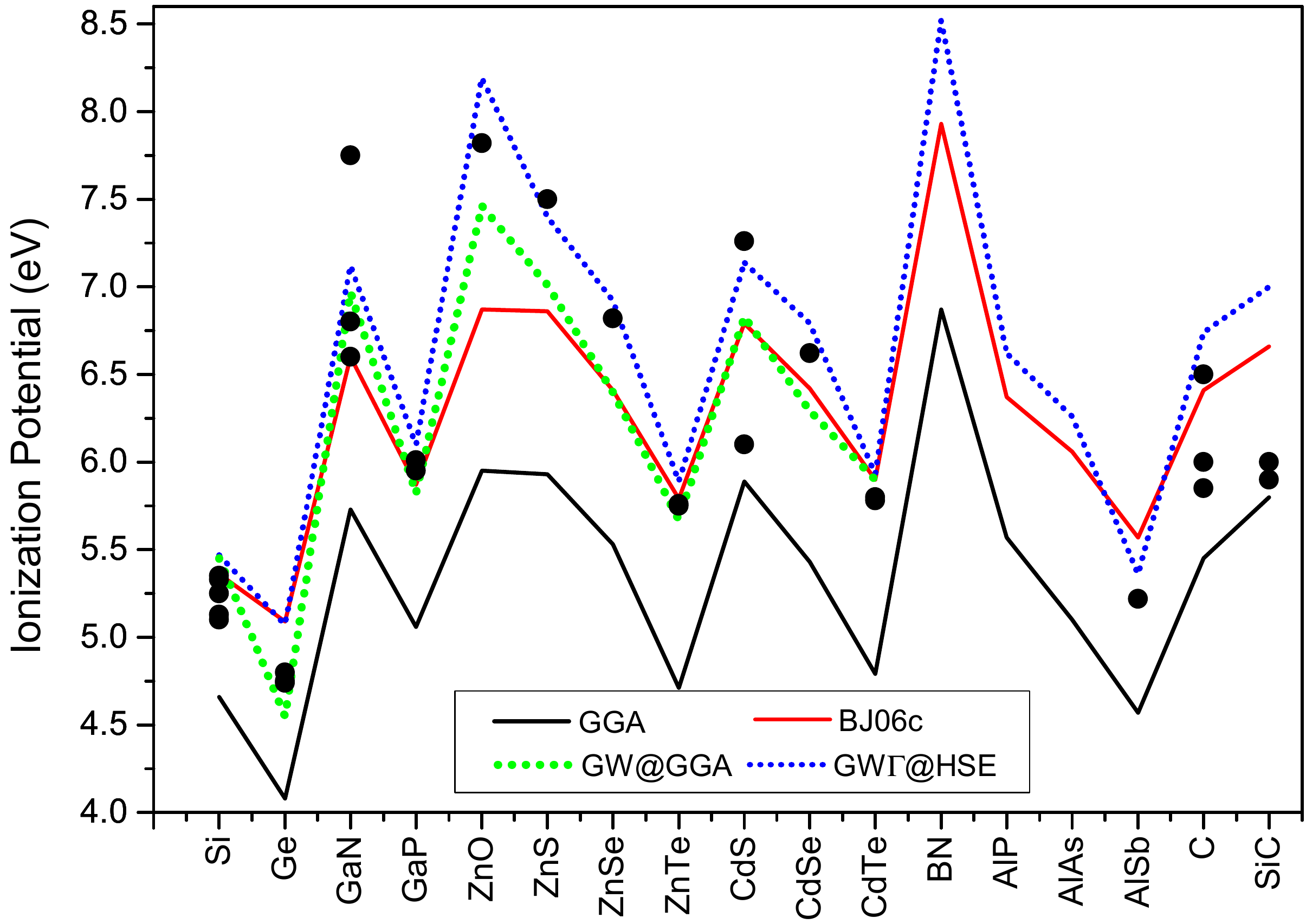}\\
\caption{(Color on line) IP's of the 17 semiconductors from GGA (black solid line) and BJ06c (red solid line) calculated by Eq.(\ref{IPsolidFinal}). The experimental data (black dots) and the GW results (blue and green dotted lines) are the same as Table \ref{IPs}.\label{main}}
\end{figure}

Using BJ06c within ASG, we have calculated the IP's of 17 semiconductors. The results are presented in Table \ref{IPs} and FIG.\ref{main} and are compared to GWA\cite{vaspip,jiangip} and experiment.  Through the two-step procedure, the accuracy of IP is determined by the surface part of the calculation which may be checked by comparing Eq.(\ref{IPsolidFinal}) with  $-\varepsilon_{\mathrm{vbm}}^{\mathrm{slab}}$, the negative eigenvalue of the VBM of the slab. This is because, providing surface states are absent, then Eq.(\ref{IPsolidFinal}) and $-\varepsilon_{\mathrm{vbm}}^{\mathrm{slab}}$  shall give the same IP value unless the slab is not sufficiently thick. Indeed, for 14 semiconductors  Eq.(\ref{IPsolidFinal}) and $-\varepsilon_{\mathrm{vbm}}^{\mathrm{slab}}$ differ by less than 0.1 eV. The exceptions are  BN, SiC, CdS for which surface states are present. Therefore, the numerical accuracy of our IP's is better than 0.1 eV.
\begin{table}[!ht]
\caption{(Unit in eV) The GGA and BJ06c IP's of the 17 semiconductors calculated from Eq.(\ref{IPsolidFinal}) and the negative of the VBM eigenvalue of the slab. The GW results are calculated by GW@GGA (Ref.\cite{jiangip}) and GW$\Gamma$@HSE (Ref.\cite{vaspip}). Experimental results are collected from these two works. For the original reports see references therein.\label{IPs}}
\begin{tabular*}{\linewidth}{l|cc|cc|cc|l}
\hline\hline
\multirow{2}{*}{} & \multicolumn{2}{c|}{\parbox[c]{1.6cm}{GGA}} & \multicolumn{2}{c|}{\parbox[c]{1.6cm}{BJ06c}} & \multicolumn{2}{c|}{\parbox[c]{1.6cm}{GW}} & \multirow{2}{*}{~experiments}\\
\cline{2-7}
& \parbox[c]{0.8cm}{Eq.(\ref{IPsolidFinal})} & \parbox[c]{0.8cm}{$\varepsilon_{\mathrm{vbm}}^{\mathrm{slab}}$}
& \parbox[c]{0.8cm}{Eq.(\ref{IPsolidFinal})} & \parbox[c]{0.8cm}{$\varepsilon_{\mathrm{vbm}}^{\mathrm{slab}}$}
& \parbox[c]{0.8cm}{Ref.\cite{jiangip}} & \parbox[c]{0.8cm}{Ref.\cite{vaspip}}  \\
\hline
Si    &  4.66 & 4.68 &   5.36 & 5.38 & 5.45  & 5.47 & 5.13, 5.33, 5.1,\\
      &       &      &        &      &       &      & ~~ 5.25, 5.35 \\
Ge    &  4.08 & 4.07 &   5.09 & 5.03 & 4.55  & 5.07 & 4.75, 4.8, 4.74 \\
GaN   &  5.73 & 5.68 &   6.60 & 6.53 & 6.97  & 7.12 & 6.6, 6.8, 7.75 \\
GaP   &  5.06 & 5.04 &   5.87 & 5.83 & 5.82  & 6.10 & 5.95, 6.01\\
ZnO   &  5.95 & 6.00 &   6.87 & 6.88 & 7.46  & 8.19 & 7.82 \\
ZnS   &  5.93 & 5.90 &   6.86 & 6.81 & 7.01  & 7.40 & 7.5 \\
ZnSe  &  5.53 & 5.50 &   6.41 & 6.37 & 6.40  & 6.92 & 6.82 \\
ZnTe  &  4.71 & 4.69 &   5.79 & 5.76 & 5.67  & 5.89 & 5.76, 5.75 \\
CdS   &  5.89 & 5.79 &   6.79 & 6.62 & 6.83  & 7.14 & 6.1, 7.26 \\
CdSe  &  5.43 & 5.38 &   6.42 & 6.33 & 6.29  & 6.79 & 6.62 \\
CdTe  &  4.79 & 4.75 &   5.90 & 5.82 & 5.90  & 5.93 & 5.78, 5.8 \\
BN    &  6.87 & 6.51 &   7.93 & 7.57 & --    & 8.52 & -- \\
AlP   &  5.57 & 5.51 &   6.37 & 6.32 & --    & 6.62 & -- \\
AlAs  &  5.10 & 5.03 &   6.06 & 5.99 & --    & 6.26 & -- \\
AlSb  &  4.57 & 4.52 &   5.57 & 5.51 & --    & 5.36 & 5.22 \\
C     &  5.45 & 5.53 &   6.41 & 6.49 & --    & 6.74 & 5.85, 6.0, 6.5 \\
SiC   &  5.80 & 5.60 &   6.66 & 6.45 & --    & 7.00 & 5.9, 6.0 \\
\hline\hline
\end{tabular*}
\end{table}

Consistent with earlier works, GGA\cite{PBE} consistently underestimates the experimental IP's by about 1 eV. With BJ06c, all IP's are upshifted, and the corrections upon GGA are from 0.70 eV (for Si) to 1.11 eV (for CdTe). Experimentally, the IP measurement is extremely sensitive to the samples' surface condition which has led to large scattering of the IP data. Consequently, strict quantitative comparison between theories and experiment is impossible. Nevertheless, the extent of the corrections by BJ06c is in the desired range and is similar to GWA. The largest discrepancy is found for ZnO\cite{ZnO} which is also a difficult case for GWA. In general, the performance of BJ06c is quite impressive since it is a local potential, requesting essentially negligible computational cost compared to GWA.

From Table \ref{IPs} and FIG.\ref{main} the performance of BJ06c is about equally good for solids as for atoms\cite{OEPip} and small molecules\cite{Gorling}. This contradicts the earlier report\cite{Engel} that OEP is not useful for semiconductor IP.  In fact, we have also calculated the IP of graphene by BJ06c and obtained 4.6 eV which is in excellent agreement with experiment. Therefore, a repulsive correlation potential is not needed. We note in Ref.\cite{Engel} the OEP was generated by supercell. Limited by the low efficiency of the supercell method, the sheet separation is only 14 Bohr which is definitely not sufficient to achieve the asymptotic value of the OEP at the vacuum edge.

The long-range part of $v_{xc}$ has been well understood for finite systems, but is widely ignored in surface calculations for which the erroneously short-ranged LDA/GGA are almost exclusively used. To check its effects for solid surface, we have decomposed $\Delta$IP, the total IP correction upon GGA,  into the three individual contributions of $\Delta\varepsilon_{\mathrm{vbm}}^{\mathrm{p.cell}}$, $\Delta \overline{v}_{\mathrm{Coul}}^{\mathrm{p.cell}}$, and $\Delta \overline{v}_{\mathrm{Coul}}^{\mathrm{s.cell}}$ following Eq.(\ref{IPsolidFinal}). With this decomposition the effects of the long-range part of the $v_{xc}$ are only reflected in the surface term since in the bulk calculation there is no asymptotic region. Note that the electron self-energy also shares similar asymptotic behavior\cite{GWasymp}, the existing GWA studies\cite{vaspip,jiangip,imprvgwa} can therefore be used as convenient examples to assist our analysis. For this purpose, let us recall that all GWA calculations are only applied to the bulk, while the slabs are always treated by GGA so that:
\begin{align}
\label{gwa}
\mathrm{IP}^{\mathrm{GWA}} = \mathrm{IP}^{\mathrm{GGA}} - \Delta\varepsilon_{\mathrm{vbm}}^{\mathrm{p.cell}}+ \Delta \overline{v}_{\mathrm{Coul}}^{\mathrm{p.cell}}.
\end{align}
In this way,  $\Delta \overline{v}_{\mathrm{Coul}}^{\mathrm{s.cell}}$, together with all effects of the long-range part of $v_{xc}$, are completely ignored. In fact, if the GWA of the bulk is performed in the one-shot fashion, i.e. without self-consistency, then  $\Delta \overline{v}_{\mathrm{Coul}}^{\mathrm{p.cell}}$ is also ignored. Since Coulomb potential is uniquely determined by density,  the underlying assumption  is that the GGA densities are already accurate enough to be directly used for the IP calculations.

\begin{table}[!ht]
\caption{(Units in eV) Total IP correction $\Delta$IP and the individual contributions of $\Delta\varepsilon_{\mathrm{vbm}}^{\mathrm{p.cell}}$, $\Delta \overline{v}_{\mathrm{Coul}}^{\mathrm{p.cell}}$, and $\Delta \overline{v}_{\mathrm{Coul}}^{\mathrm{s.cell}}$ from Eq.(\ref{IPsolidFinal}). PCT gives the percentages of the contributions.\label{correction}}
\begin{tabular*}{\linewidth}{l|c|cc|cc|cc}
\hline\hline
  \multirow{2}{*}{}
& \multirow{2}{*}{\parbox[c]{1.cm}{$\Delta$IP}}
& \multicolumn{2}{c|}{\parbox[c]{1.9cm}{$\Delta \varepsilon_{\mathrm{vbm}}^{\mathrm{p.cell}}$}}
& \multicolumn{2}{c|}{\parbox[c]{1.9cm}{$\Delta \overline{v}_{\mathrm{Coul}}^{\mathrm{p.cell}}$}}
& \multicolumn{2}{c}{\parbox[c]{1.9cm}{$\Delta \overline{v}_{\mathrm{Coul}}^{\mathrm{s.cell}}$}} \\
\cline{3-8}
&
& \parbox[c]{0.95cm}{value} & \parbox[c]{0.95cm}{PCT}
& \parbox[c]{0.95cm}{value} & \parbox[c]{0.95cm}{PCT}
& \parbox[c]{0.95cm}{value} & \parbox[c]{0.95cm}{PCT}  \\
\hline
Si       & 0.70   &   -0.62 & 89  &   -0.11 & 16 &  -0.03 & -4\\
Ge 	     & 1.00   &   -0.88 & 88  &   -0.20 & 20 &  -0.08 & -8\\
GaN      & 0.87   &   -0.78 & 90  &   -0.11 & 13 &  -0.02 & -2\\
GaP      & 0.80   &   -0.56 & 70  &   -0.27 & 34 &  -0.02 & -3\\
ZnO      & 0.93   &   -1.07 & 115 &    0.11 &-12 &  -0.03 & -3\\
ZnS      & 0.93   &   -0.73 & 78  &   -0.22 & 24 &  -0.02 & -2\\
ZnSe     & 0.88   &   -0.74 & 84  &   -0.18 & 20 &  -0.04 & -5\\
ZnTe     & 1.07   &   -0.75 & 70  &   -0.45 & 42 &  -0.13 & -12\\
CdS      & 0.90   &   -0.75 & 83  &   -0.21 & 23 &  -0.06 & -7\\
CdSe     & 1.00   &   -0.76 & 76  &   -0.30 & 30 &  -0.07 & -7\\
CdTe     & 1.11   &   -0.78 & 70  &   -0.47 & 42 &  -0.15 & -14\\
BN 	     & 1.06   &   -0.83 & 78  &   -0.22 & 21 &   0.01 & 1\\
AlP      & 0.80   &   -0.64 & 80  &   -0.18 & 23 &  -0.02 & -3\\
AlAs     & 0.96   &   -0.72 & 75  &   -0.28 & 29 &  -0.04 & -4\\
AlSb     & 1.00   &   -0.80 & 80  &   -0.33 & 33 &  -0.12 & -12\\
C        & 0.96   &   -0.73 & 76  &   -0.22 & 23 &   0.01 & 1\\
SiC      & 0.87   &   -0.72 & 83  &   -0.16 & 18 &  -0.01 & -1\\
\hline\hline
\end{tabular*}
\end{table}

From Table \ref{correction}, the average contribution of the three terms to the total $\Delta$IP are 82\%, 23\% and -5\%, respectively. Therefore, the bulk part of the corrections indeed accounts for most of the $\Delta$IP which lends fundamental support to the GWA treatment. Especially, the smallness of  $\Delta \overline{v}_{\mathrm{Coul}}^{\mathrm{p.cell}}$ implies that the GGA bulk density is indeed of high quality. Problem, however, exists in  $\Delta\overline{v}_{\mathrm{Coul}}^{\mathrm{s.cell}}$ since its 23\% contribution to the total $\Delta$IP is certainly non-negligible. Especially, for the two Te compounds the contributions are larger than 40\%. This means that the GGA slab density is not sufficiently accurate.

The reason why the two GGA densities are of different quality is because SI behaves differently in the bulk and the slab: Within the bulk, SI usually causes the density to over-delocalize. However, this problem mostly affects strongly correlated materials, while for weak to intermediate correlation it is not very serious. In fact, within the bulk the SI error is largely screened out by the response of Coulomb potential as having been illustrated by Li {\it et al.}\cite{KLI2}. On the other hand, in the slab case Coulomb potential cannot compensate the missing tail of GGA since it decays even faster than the $v_{xc}$. Consequently, the SI error is more prevailing in the slab than in the bulk.

\begin{figure}[h]
\centering
\includegraphics[width=\linewidth]{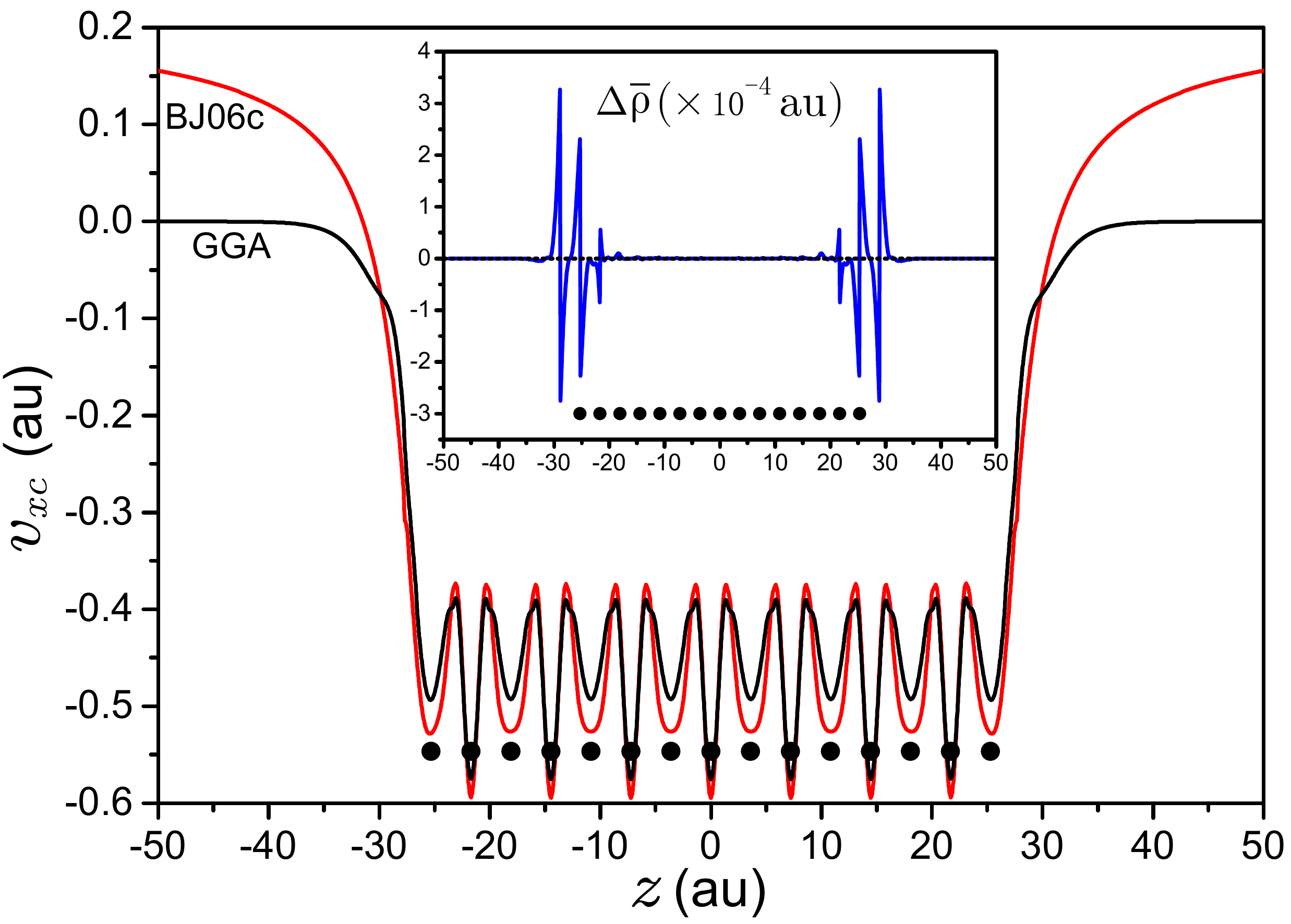}
\caption{\label{vxc2} (Color on line) $v_{xc}$ of the GGA (black) and BJ06c (red) for the ZnS slab of FIG.\ref{ZnSslab} along $(0,0,z)$. The plot path is between atomic sites to avoid the divergence of the GGA at the nuclei. Inset shows the difference of the macroscopically averaged\cite{Fall} slab  density $\Delta\bar{\rho}(z)=\bar{\rho}^{\mathrm{BJ06c}}(z)-\bar{\rho}^{\mathrm{GGA}}(z)$. Black dots indicate the location of the atomic planes.}
\end{figure}

To make sure that $\Delta \overline{v}_{\mathrm{Coul}}^{\mathrm{s.cell}}$ is indeed related to the long-range part of BJ06c, we have plotted the two $v_{xc}$'s for FIG.\ref{ZnSslab} together with the difference of their macroscopically averaged\cite{Fall} density.  FIG.\ref{vxc2} shows that within the bulk BJ06c is deeper than GGA which explains the downshift of $\varepsilon_{\mathrm{vbm}}$. The most obvious distinction, however, is around the slab surface: While GGA already decays to zero at about $z=40$, BJ06c climbs up very slowly to its 0.20 value at the vacuum edge of $z=247$ (not shown). Correspondingly, within the bulk the electron density is essentially unchanged. Only near the surface does the slow variation of BJ06c strongly perturb the surface density. For the relaxed slab, this enhances the surface dipole and changes $\overline{v}_{\mathrm{Coul}}^{\mathrm{s.cell}}$ by -0.22 eV which contributes positively to the total $\Delta$IP of 0.93 eV.

Compared to atoms and molecules, the effects of the long-range part of the $v_{xc}$ are weaker for solid surface. This is because the wave functions are mainly bound to the slab while less exposed to the vacuum or the asymptotic region. At this point, it is necessary to point out that the exact asymptotic behavior of $v_{xc}$ for solid surface is still unresolved\cite{mypaper,Engel}. In BJ06c, the asymptotic behavior $-1/z+C$ of the exchange part is exact according to the form found in Ref.\cite{theorem2}, since the $C$ constant has no effect on the slab density or $\Delta \overline{v}_{\mathrm{Coul}}^{\mathrm{s.cell}}$. On the other hand, the LDA correlation is still erroneously short-ranged. Consequently, the $\Delta \overline{v}_{\mathrm{Coul}}^{\mathrm{s.cell}}$ data in Table \ref{correction} contain systematic errors which are hard to evaluate. Nevertheless, $\Delta \overline{v}_{\mathrm{Coul}}^{\mathrm{s.cell}}$ is at least system dependent, and there is no apparent reason why it can  be universally neglected.  For an example, it is found that GGA can successfully lineup the Coulomb potential at the interface of hetero-structures. This is frequently cited to support the similar use of GGA for the potential alignment in the IP calculation by GWA\cite{jiangip,vaspip,imprvgwa}.  However,  the two interfaces are fundamentally different: The asymptotic behavior of $v_{xc}$ is operative only at the slab-vacuum interface,  but has little effect at the interface of hetero-structure because, again, there is no asymptotic region.

Indirect evidence supporting non-negligible $\Delta\overline{v}_{\mathrm{Coul}}^{\mathrm{s.cell}}$ may be found in  existing GWA results. As Table \ref{correction} shows, except for ZnO, all  $\Delta\overline{v}_{\mathrm{Coul}}^{\mathrm{s.cell}}$ contribute positively to $\Delta$IP. Therefore, by neglecting this term GWA shall systematically underestimate the semiconductor IP's. Indeed, it has been consistently found\cite{jiangip,vaspip,imprvgwa} that semiconductor IP's calculated by GWA starting from GGA are systematically too low. Nevertheless, since the same GWA procedure also underestimates the band gaps, it is envisioned that both problems are due to the same origin of insufficient quasiparticle corrections.

If $\Delta\overline{v}_{\mathrm{Coul}}^{\mathrm{s.cell}}$ is indeed universally negligible, then by merely improving GWA it shall be possible to achieve accurate band gaps and accurate IP's at the same time. To enhance quasiparticle corrections, hybrid functionals have been used to replace the GGA starting point\citep{vaspip,imprvgwa}. As they seem to over-correct the problems, the fraction of the Fock exchange is changed to an adjustable parameter\citep{imprvgwa}. Alternatively, vertex function from certain cross diagrams is included to scale down the corrections\citep{vaspip}. Besides, quasiparticle self-consistency\citep{qscgw} with the screened interaction of GGA\citep{imprvgwa} is also attempted. It is found, however, when the IP's seem to agree to experiment, the band gaps are unevenly overestimated\cite{vaspip}. On the other hand, when the band gaps  seem to agree to experiment, half of the IP's are overestimated\cite{imprvgwa}. Although in both cases it is claimed that further vertex corrections shall reconcile the conflicts, it is not clear how the extra corrections only affect either the IP's or the band gaps, while leaving the other set unchanged.

This work is supported by National Science Foundation of China (Grant Nos. 61390504, 91221202), and by National Basic Research Program of China (Grant Nos. 2012CB932700, 2012CB932703).


\begin{thebibliography}{99}
\bibitem{ipnote} Strictly speaking, for infinite solids the IP theorem does not hold anymore. This is because the existence of this theorem for finite systems is based on the fact that the highest occupied orbital is also the most extended orbital in space. For solid surface, however, the most extended orbital is usually the highest occupied orbital at the $\Gamma$-point of the planar Brillouin zone which may not be the VBM (See Ref.\cite{Engel} for details).  Nevertheless, possible deviation from the IP theorem is likely small and is not considered in this work.
\bibitem{GGAnote} The choice of LDA over GGA is purely technical. Since SI typically behaves similarly in LDA and GGA, we expect similar performance for the IP's if BJ06c is amended by a GGA correlation.
\bibitem{ZnO} Albeit being an insulator, ZnO is well known for its persistent and mysterious $n$-type conductivity. It is possible that the IP measurement suffers from strong surface charging effect.
\end{thebibliography}
\end{document}